\documentclass[%
 reprint,
superscriptaddress,
 amsmath,amssymb,
 aps,
]{revtex4-2}

\usepackage{graphicx}
\usepackage{dcolumn}
\usepackage{bm}
\usepackage{hyperref}

\begin{document}

\preprint{APS/123-QED}

\title{Unveiling quasiparticle dynamics of topological insulators through Bayesian modelling}

\author{Satoru Tokuda}
\email{s.tokuda.a96@m.kyushu-u.ac.jp}
\affiliation{Research Institute for Information Technology, Kyushu University, Kasuga 816-8580, Japan}
\affiliation{Mathematics for Advanced Materials-OIL, AIST, Sendai 980-8577, Japan}

\author{Seigo Souma}
\affiliation{Center for Spintronics Research Network, Tohoku University, Sendai 980-8577, Japan.}
\affiliation{Advanced Institute for Materials Research (WPI-AIMR), Tohoku University, Sendai 980-8577, Japan.}

\author{Kouji Segawa}
\affiliation{Department of Physics, Kyoto Sangyo University, Kyoto 60 3-8555, Japan.}

\author{Takashi Takahashi}
\affiliation{Center for Spintronics Research Network, Tohoku University, Sendai 980-8577, Japan.}
\affiliation{Advanced Institute for Materials Research (WPI-AIMR), Tohoku University, Sendai 980-8577, Japan.}
\affiliation{Department of Physics, Tohoku University, Sendai 980-8578, Japan.}

\author{Yoichi Ando}
\affiliation{Institute of Physics II, University of Cologne, K\"{o}ln 50937, Germany.}

\author{Takeshi Nakanishi}
\affiliation{Mathematics for Advanced Materials-OIL, AIST, Sendai 980-8577, Japan}

\author{Takafumi Sato}
\email{t-sato@arpes.phys.tohoku.ac.jp}
\affiliation{Center for Spintronics Research Network, Tohoku University, Sendai 980-8577, Japan.}
\affiliation{Advanced Institute for Materials Research (WPI-AIMR), Tohoku University, Sendai 980-8577, Japan.}
\affiliation{Department of Physics, Tohoku University, Sendai 980-8578, Japan.}

\date{\today}

\begin{abstract}
Quasiparticle - a key concept to describe interacting particles - characterizes electron-electron interaction in metals (Fermi liquid) and electron pairing in superconductors. While this concept essentially relies on the simplification of hard-to-solve many-body problem into one-particle picture and residual effects, a difficulty in disentangling many-body effects from experimental quasiparticle signature sometimes hinders unveiling intrinsic low-energy dynamics, as highlighted by the fierce controversy on the origin of Dirac-band anomaly in graphene and dispersion kink in high-temperature superconductors. Here, we propose an approach to solve this fundamental problem - the Bayesian modelling of quasiparticles. We have chosen a topological insulator $\mathrm{TlBi(S,Se)_2}$ as a model system to formulate an inverse problem of quasiparticle spectra with semiparametric Bayesian analysis, and successfully extracted one-particle and many-body characteristics, i.e. the intrinsic energy gap and unusual lifetime in Dirac-quasiparticle bands. Our approach is widely applicable to clarify the quasiparticle dynamics of quantum materials.
\end{abstract}

\maketitle


Low-energy excitations in interacting electronic systems are known to be characterized by quasiparticles. The concept of quasiparticle was originally proposed in the Laudau’s Fermi-liquid theory wherein strongly interacting electrons share a similar behavior with weakly interacting counterparts. Instead of strongly interacting bare electrons (holes), one can define dressed electrons (holes) as elementary excitations (i.e., quasiparticles), which can be understood by extending the framework of single-particle approximation. While the Fermi-liquid theory successfully captured low-energy dynamics of normal metals and $^3$He, the quasiparticle concept is nowadays applied widely in solids such as electron systems interacting with lattice vibrations (phonons) and spin excitations (magnons). Angle-resolved photoemission spectroscopy (ARPES) has played a pivotal role in uncovering key quasiparticle properties by capturing the energy dispersion ($E$-$k$ relation) and lifetime of e.g., Bogoliubov quasiparticles associated with the superconducting Cooper pairing in high-temperature superconductors \cite{bogoljubov1958new, campuzano1996direct, matsui2003bcs} and mass-renormalized quasiparticles caused by strong electron-phonon coupling on metal surfaces and quasi-two-dimensional materials \cite{valla1999many, hengsberger1999photoemission, lanzara2001evidence}. As highlighted by these examples, for the understanding of the origin and mechanism of exotic physical properties of novel materials, it is crucial to experimentally establish the nature of quasiparticles.

To elucidate the quasiparticle dynamics, it is desirable to be able to unambiguously extract the original single-particle band dispersion (bare-band dispersion, $E_k$) and many-body effects (self-energy, $\Sigma$) from the ARPES data. Both of these physical quantities are directly linked to the ARPES spectrum through the spectral function expressed as,
\begin{align}
A(k, \omega)=\frac{1}{\pi} \frac{-\operatorname{Im} \sum(\mathbf{k}, \omega)}{\left[\omega-E_{k}-\operatorname{Re} \sum(\mathbf{k}, \omega)\right]^{2}+\left[\operatorname{Im} \sum(\mathbf{k}, \omega)\right]^{2}}
\end{align}
where $\omega$ is the energy with respect to the Fermi level ($E_F$). Many attempts have been hitherto made to extract the intrinsic $\Sigma$ from ARPES data by assuming a reasonable shape of $E_k$. For example, $E_k$ was referenced to the band calculation obtained with the local density approximation (e.g., \cite{bogdanov2000evidence, meevasana2008extracting}), or it was empirically approximated with a polynomial function (e.g., linear or parabola \cite{johnson2001doping, kordyuk2005bare}). While such data analysis certainly gave insights into the quasiparticle dynamics, one often faced a serious problem in clarifying the nature of many-body interactions. This is represented by the fierce debates on the absence or appearance of an intrinsic energy gap at the Dirac point in epitaxial single-layer graphene \cite{bostwick2010observation, zhou2007substrate} which is critical for feasible application of graphene as a semiconductor device. Also, the origin of dispersion kink in cuprate superconductors (phononic, magnetic, or others) is controversial for more than a decade \cite{damascelli2003angle}, and its relationship with the high-$T_c$ mechanism is yet to be clarified. These controversies partially originate from a few assumptions one had to make to extract $E_k$ and $\Sigma$. 

To overcome these problems, we apply semiparametric Bayesian modelling to ARPES data. We start by introducing the basics of Bayesian analysis through a simple demonstration. As a prototypical example, we model the ARPES intensity of a topological insulator (TI) $\mathrm{TlBi(S,Se)_2}$ \cite{sato2011unexpected} based on the parametric form of bare band dispersion and nonparametric forms of any other elements to perform the semiparametric Bayesian analysis of spectral function. We provide a clear insight into one-particle and many-body characteristics of $\mathrm{TlBi(S,Se)_2}$ by successfully extracting bare-band dispersion and self-energy.

\section*{Results}
\begin{figure}[tbp]
\begin{center}
\begin{tabular}{c}
\includegraphics[width=8.6cm]{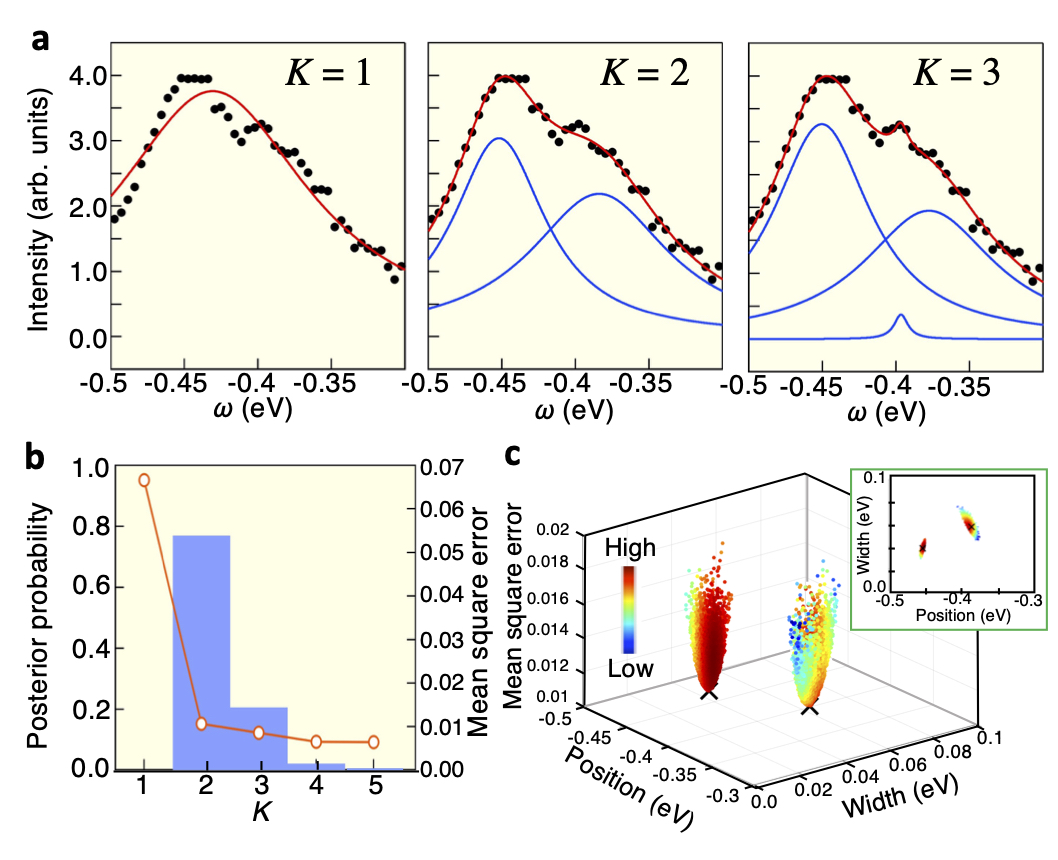}
\end{tabular}
\end{center}
\caption{\textbf{Application of Bayesian analysis to the energy distribution curve.} {\bf a}, Numerical fittings (red curves) to a typical energy distribution curve (EDC; black dots) assuming the existence of single, double, or triple intrinsic Lorentzian peaks (blue curves). A hyper-parameter $K$ represents the number of peaks. {\bf b}, Plot of mean square error (red open circles) and posterior probability (blue rectangles) against $K$ that supports the existence of two peaks in the EDC. {\bf c}, Density scatter plot of the posterior probability distribution in the Bayesian analysis (color dots), compared with the least-squares solution (black crosses). Inset shows the posterior probability distribution as a projection in the direction of mean square error.}
\label{fig:Fig.1}
\end{figure}

\noindent
\textbf{Basics of Bayesian analysis.}
First, we explain the basic concept of Bayesian analysis, by showing its application to an energy distribution curve (EDC) contributed by multiple bands. A common approach in extracting the peak positions (i.e., contributing energy bands) is to find a good reproduction of the experimental EDCs by simulated EDCs using the least-square method. However, this method cannot pin down which class of model (e.g., how many bands are contributing) is the best for given data, often posing a question on the basic applicability of the model itself. To demonstrate this problem, we show in Fig. \ref{fig:Fig.1}a a representative experimental EDC (dots) together with the result of numerical fittings (red curves) using the least-square method assuming the existence of intrinsic single, double, and triple Lorentzian peaks (blue curves; the number of peaks $K = 1$-$3$) that represent three different class of models. One can immediately recognize that the model with triple Lorentzian peaks shows the best fit to the experimental data. This is natural because the inclusion of more peaks (more parameters) always leads to a decrease in mean square error (shown by red line in Fig. \ref{fig:Fig.1}b). However, it does not validate that the actual number of peaks are three. One needs to select the most appropriate model (in this case, the number of peaks) to suitably reproduce the EDC. Importantly, such selection should not include arbitrariness and must not rely on “human eyes”. The Bayesian framework, as an extension of the least squares, enables the evaluation of the model’s appropriateness itself in terms of the posterior probability derived from the chain rule of probability, called Bayes’ formula (see Methods). One can see from Fig. \ref{fig:Fig.1}b that the double-peak model ($K = 2$) has the highest probability (77 \%) among $K = 1$-$5$. Whereas the least-square method determines a unique solution of the parameter set as a global minimum of the mean square error, the Bayesian framework treats the “statistical ensemble” of numerous solutions as random variables with the Boltzmann distribution, called the posterior probability distribution of the parameter set. Namely, many solutions for peak position and peak width are plotted in the parameter space and colored with the posterior probability density proportional to the Boltzmann factor, where the point with the highest posterior probability density corresponds to the best-fit parameters (least-square solution) as highlighted in Fig. \ref{fig:Fig.1}c [note that ‘marginal’ posterior probability density is plotted in Fig. \ref{fig:Fig.1}c, because the intensity of each peak (not shown) is also a model parameter]. Like the canonical ensemble, a statistical ensemble of Bayesian analysis is conditioned on three principal factors: the number of observed data points (e.g., the number of meshes in an ARPES image), the number of parameters in each model class, and the “temperature”. This temperature is nothing to do with actual temperature of the sample and is related to the signal-to-noise ratio of observed data, affecting the uncertainty of model parameters. If the “temperature” is treated as a hyper-parameter, it can be estimated by maximizing the “partition function”, which is connected with the posterior probability of model class (Fig. \ref{fig:Fig.1}b) through Bayes’ formula (see Methods).

\begin{figure}[tbp]
\begin{center}
\begin{tabular}{c}
\includegraphics[width=8.6cm]{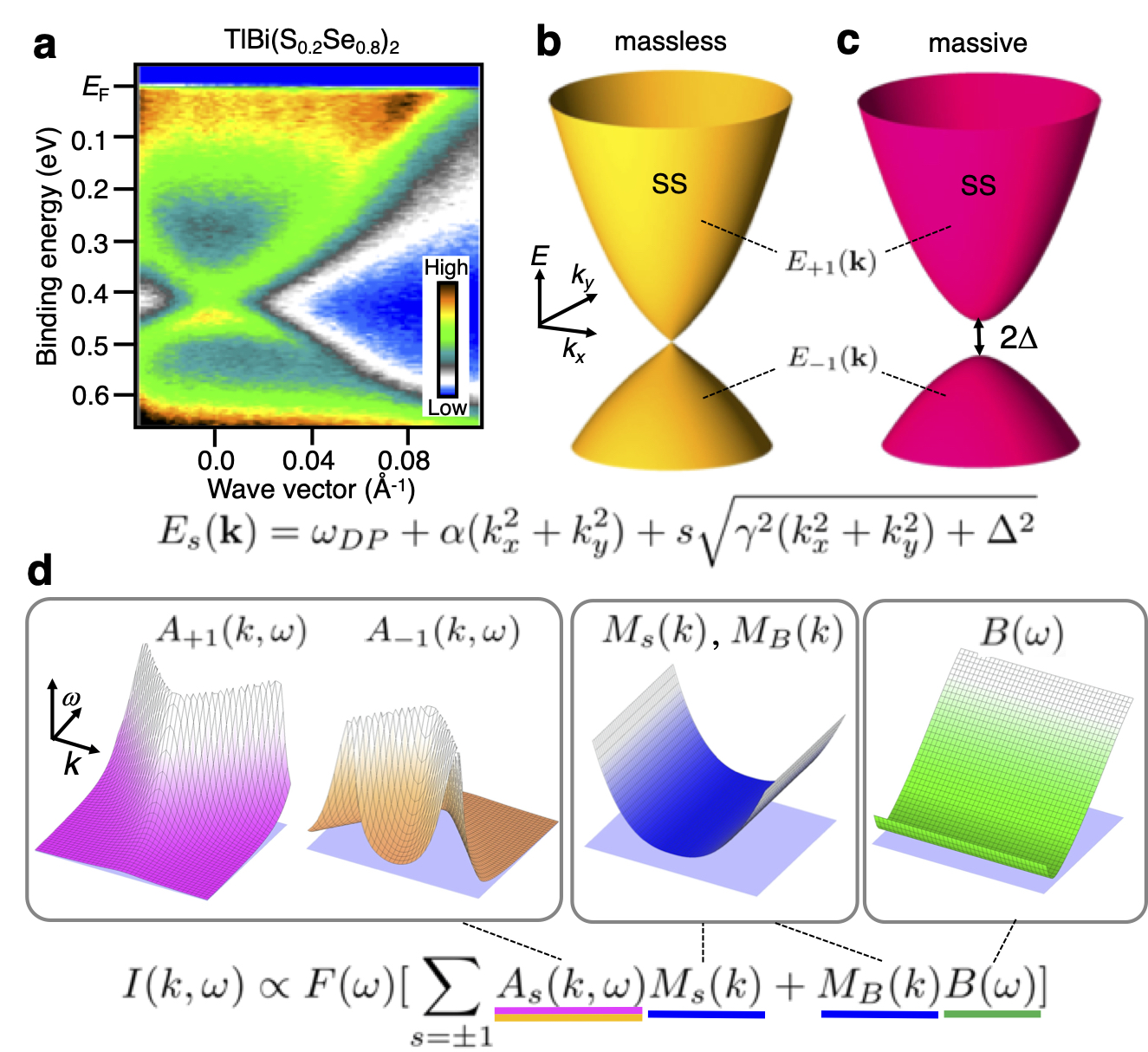}
\end{tabular}
\end{center}
\caption{\textbf{Semiparametric Bayesian modelling of the gapped Dirac-cone surface state.} {\bf a}, Angle-resolved photoemission spectroscopy (ARPES) intensity plot in the vicinity of the Fermi level $E_F$ around the $\Gamma$ point for $\mathrm{TlBi(S_{0.2}Se_{0.8})_2}$ measured at $T$ = 30 K with the Xe-I$\alpha$ line ($h\nu = 8.437$ eV) \cite{sato2011unexpected}. {\bf b}, {\bf c}, Schematic energy dispersion of gapless and gapped Dirac-cone bands, respectively. The full energy gap is $2\Delta$. Analytical form of the bare-band dispersion $E_s(k)$ used in the model is also shown at the bottom; $s$, $\omega_{DP}$, $\alpha$, $\gamma$, $\Delta$ are branch index [upper ($s = +1$) or lower ($s = -1$) Dirac cone], Dirac-point energy, parabolic dispersion term, Dirac velocity, and Dirac gap, respectively. {\bf d}, Example of simulated ARPES intensity $I(k, \omega)$ used in the semiparametric Bayesian analysis, which is composed of photoelectron matrix-element term for the surface band $M_s(k)$, single-particle spectral function $A_s(k, \omega)$, photoelectron matrix-element term for the background $M_B(k)$, and angle-integrated-type background $B(\omega)$. $F(\omega)$ denotes the Fermi-Dirac distribution function.}
\label{fig:Fig.2}
\end{figure}

\noindent
\textbf{Semiparametric model of ARPES intensity.}
Now that the basic scheme of the Bayesian analysis is demonstrated for a single spectrum, we apply this analysis to the actual two-dimensional (2D) ARPES intensity through a semiparametric modelling of quasiparticle bands. For a testbed system, we have chosen $\mathrm{TlBi}(\mathrm{S}_{1-x} \mathrm{Se}_x)_2$ ($x = 0.8$) \cite{sato2011unexpected}, where a slight intensity suppression around the Dirac point is seen as shown in Fig. \ref{fig:Fig.2}a, but the origin of such unusual gap-like behavior has been a target of intensive debates \cite{xu2011topological, souma2012spin, li2013phonon, habe2013gapped, sanchez2016nonmagnetic, brahlek2016disorder, tanaka2018influence, wang2017excitonic, zhang2018topological, qi2019dephasing}. Since the Dirac gap is a prerequisite for realizing some exotic topological quantum phenomena \cite{hasan2010colloquium, qi2011topological, ando2013topological}, it is important to establish whether the bare band $E_k$ of the Dirac-cone state is gapless (Fig. \ref{fig:Fig.2}b) or gapped (Fig. \ref{fig:Fig.2}c) to pin down the mechanism of unusual Dirac-band anomaly (it is noted here, if the chemical potential is situated within the gap, physical properties are mainly governed by the “quasiparticle band gap” which is a combination of the bare-band gap and the self-energy effect). It is worth noting that such absence or appearance of the Dirac gap is also critical for the classification of magnetic-TI and axion-insulator phases, as highlighted by a fierce debate on whether or not the surface state hosts the Dirac gap associated with the time-reversal-symmetry-breaking magnetic order in $\mathrm{MnBi_2Te_4}$ and related compounds (see e.g. \cite{li2019dirac, chen2019intrinsic, gong2019experimental, otrokov2019prediction, hao2019gapless, chen2019topological}). In our Bayesian analysis, we treat such distinct gapless and gapped states as different model classes and judge the validity of the models for given ARPES data. As shown in the bottom of Figs. \ref{fig:Fig.2}a-c, we assume that $E_k$ is represented by the function $E_s(k)$ for the band index $s = \pm 1$ ($+1$ for upper, $-1$ for lower Dirac cones), parametrized by the binding energy ($E_B$) at the Dirac point $\omega_{DP}$, the band asymmetry $\alpha$ (this parameter is associated with the effective-mass asymmetry of bulk conduction and valence bands \cite{zhang2009topological, lu2010massive}), a band parameter $\gamma$, and the half-width of band gap $\Delta$ ($\Delta=0$ for gapless, $\Delta>0$ for gapped). Our first goal is to extract the actual single-particle spectral function $A_s(k, \omega)$ from the ARPES data by estimating a parameter set of $\{\omega_{DP}, \alpha, \gamma, \Delta\}$ ($\{\omega_{DP}, \alpha, \gamma\}$ for $\Delta= 0$ case) and also obtain a concrete form of self-energy $\Sigma(k, \omega)$ (for simplicity, we assume that $\Sigma$ is $k$-independent because the $k$ range of interest is sufficiently small).

As highlighted in Fig. \ref{fig:Fig.2}d, the ARPES intensity is composed not only of intrinsic spectral function $A_s(k, \omega)$ for the Dirac-cone states, but also of the spectral weight from other features such as bulk states and spectral background. Here, all these “background” states are represented by a single $\omega$-dependent function $B(\omega)$. We also assumed independent matrix-elements $M_s(k)$ and $M_B(k)$ for the photoelectron intensities \cite{damascelli2003angle} for the Dirac-cone and the background, respectively, and simulated the total ARPES intensity $I(k, \omega)$ by neglecting the instrumental resolution. Noticeably, it is not necessary to assume any particular analytical forms of $\Sigma(\omega)$, $B(\omega)$, $M_s(k)$, and $M_B(k)$ (note that $\Sigma$ is chosen to satisfy the Kramers-Kronig relation under assumption of the particle-hole symmetry), and we could approximately treat them as vectors whose elements are values of each function at every observed points $(\omega, k)$, such as $\{\mathrm{Im} \Sigma(\omega_1), \mathrm{Im} \Sigma(\omega_2), \cdots, \mathrm{Im} \Sigma(\omega_n)\}$ at $\{\omega_1, \omega_2, \cdots , \omega_n\}$.

\noindent
\textbf{Semiparametric Bayesian analysis of ${\bf TlBi(S,Se)_2}$.}
Based on the above modelling of an ARPES image, we formulate the semiparametric Bayesian analysis of the overall spectral quantities, $E_s(k)$, $\Sigma(\omega)$, $B(\omega)$, $M_s(k)$, and $M_B(k)$ and implement this analysis by a basic algorithm for Bayesian analyses (see Methods). First, we have validated our methodology by a demonstration using mimic ARPES images that the electronic structures are predefined as ground truths (see Supplementary Note 1 and Fig. \ref{fig:Fig.S1}). Then, we have applied our methodology to the actual ARPES image containing sufficiently fine $E$-$k$ mesh (111×111 for Fig. \ref{fig:Fig.2}a) and have succeeded in simultaneously estimating all the spectral quantities [specifically, 559 (558) scalar variables for the gapped (gapless) case].

\begin{figure}[tbp]
\begin{center}
\begin{tabular}{c}
\includegraphics[width=8.6cm]{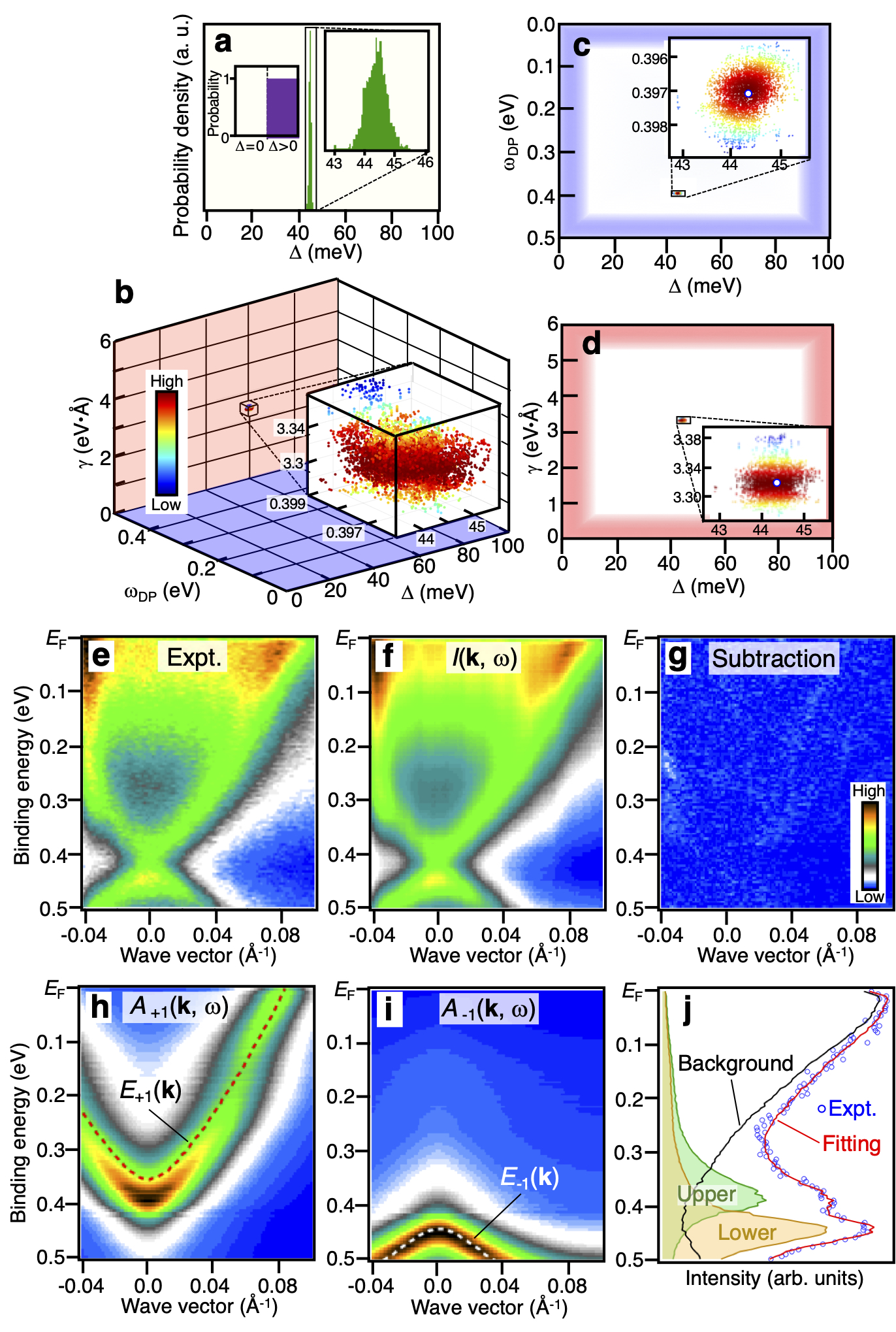}
\end{tabular}
\end{center}
\caption{\textbf{Extraction of essential band parameters from the semiparametric Bayesian analysis.} {\bf a}, Distribution of posterior probability against $\Delta$ obtained from the semiparametric fittings of the experimental angle-resolved photoemission spectroscopy (ARPES) intensity in Fig. \ref{fig:Fig.2}a with theoretical ARPES intensity $I(k, \omega)$. Left inset shows the plot of posterior probability for $\Delta = 0$ and $\Delta > 0$, where $\Delta$ denotes half-width of band gap. Right inset shows the magnified view of posterior probability against $\Delta$ around the peak. {\bf b}, 3D contour plot of posterior probability as a function of $\Delta$, the binding energy at the Dirac point $\omega_{DP}$, and the Dirac velocity $\gamma$. {\bf c}, {\bf d}, Intensity maps of posterior probability in the $(\Delta, \omega_{DP})$ and $(\Delta, \gamma)$ space, respectively. {\bf e}, ARPES-intensity plot of $\mathrm{TlBi(S_{0.2}Se_{0.8})_2}$ (same as Fig. \ref{fig:Fig.2}a). {\bf f}, Reproduced $I(k, \omega)$ obtained from the semiparametric fittings to the experimental data. {\bf g}, Subtraction of e and f. {\bf h}, {\bf i} Extracted single-particle spectral function $A_s(k, \omega)$ for the upper and lower Dirac cones. Bare bands $E_{\pm 1}(k)$ are indicated by dashed curves. {\bf j}, Experimental EDC at the $\Gamma$ point (blue open circles) and corresponding simulated EDC (red solid curve) which is decomposed into the upper (green curve) and lower (orange curve) Dirac cones as well as spectral background $B(\omega)$ (black curve).}
\label{fig:Fig.3}
\end{figure}

Now we examine whether the Dirac-cone state is gapped or not, by using gapless ($\Delta=0$) and gapped ($\Delta>0$) models that take into account all the above spectral contributions in the Bayesian analysis. Such examination is an obvious advantage of the Bayesian analysis, and can hardly be carried out by the standard least-square method. One can immediately recognize in the left inset of Fig. \ref{fig:Fig.3}a that the probability for $\Delta=0$ is negligibly small as opposed to the case for $\Delta>0$ ($100$ \% within computational uncertainty), indicating that the Dirac gap is indeed realized in $\mathrm{TlBi(S_{0.2}Se_{0.8})_2}$, consistent with the previous study \cite{sato2011unexpected}. Then, we estimated the posterior probability distribution of $\Delta$ for the gapped models, as shown by histogram in Fig. \ref{fig:Fig.3}a, where the vertical axis corresponds to the probability density. As can be seen, estimated $\Delta$ values are sharply distributed at $44.3 \pm 0.3$ meV, as better visualized in the magnified view in the right inset. This suggests that the energy gap can be estimated with higher accuracy and reliability through our Bayesian analysis compared to the EDC analysis applied thus far to $\mathrm{TlBi}(\mathrm{S}_{1-x} \mathrm{Se}_x)_2$ \cite{sato2011unexpected, souma2012spin}.

The histogram for $\Delta$ shown in Fig. \ref{fig:Fig.3}a is obtained by integrating all the other (558) parameters so that the posterior probability distribution for the other parameters cannot be seen from the plot. We show in Fig. \ref{fig:Fig.3}b the posterior probability distribution against three essential band parameters $\omega_{DP}$, $\Delta$, and $\gamma$ in the 3D density scatter plot (note that the other 556 parameters are integrated out). One can see that the data points are sharply focused in the narrow region of the $(\omega_{DP}, \Delta, \gamma)$ parameter space. This is also visualized by plotting the distribution against two parameter sets, i.e., $\Delta$ and $\omega_{DP}$ in Fig. \ref{fig:Fig.3}c ($\Delta$ and $\gamma$ in Fig. \ref{fig:Fig.3}d) in the 2D density scatter plots which are obtained by integrating $\gamma$ ($\Delta$). From these results, we obtain $(\omega_{DP}, \Delta, \gamma) = (0.397 \pm 0.001$ eV, $44.3 \pm 0.3$ meV, $3.31 \pm 0.02$ eV$\cdot \mathrm{\AA})$ as the mean and standard deviation of distributed parameter sets. This demonstrates that the Bayesian analysis is useful not only to estimate the intrinsic band parameters from ARPES data, but also to see a correlation between different band parameters; these characteristics can hardly be obtained by the conventional data analysis.

One can further confirm the validity of the band model used in our Bayesian analysis by seeing that the experimental data are very well reproduced numerically. A side-by-side comparison of the experimental ARPES image and the numerical semiparametric regression function $I(k, \omega)$ (the mean values of band parameters are used) in Figs. \ref{fig:Fig.3}e and \ref{fig:Fig.3}f signifies the almost identical intensity distribution except for a higher noise level in the experiment. Such a good matching is highlighted by the obviously weak and featureless subtracted intensity in Fig. \ref{fig:Fig.3}g. Because all the parameters are obtained from our Bayesian analysis, now we are able to show any of $A_s(k, \omega)$, $M_s(k)$, $M_B(k)$, $B(\omega)$, and $\Sigma(\omega)$ by a 2D intensity image (see Supplementary Note 2 and Fig. \ref{fig:Fig.S2}). As an example, we show in Figs. \ref{fig:Fig.3}h and \ref{fig:Fig.3}i spectral functions for the upper and lower Dirac cones $A_{+1}(k, \omega)$ and $A_{-1}(k, \omega)$, independently. The result signifies that the dispersion of each Dirac-cone branch is rounded around the Dirac point due to the Dirac-gap opening.

To highlight the degree of agreement between the experiment and Bayesian modelling, we plot in Fig. \ref{fig:Fig.3}j the experimental EDC at the $\Gamma$ point (blue open circles) together with the numerically fitted EDC (red solid curve) that includes the background $B(\omega)$ besides the peaks from the upper and lower Dirac cones. One can see that the experimental EDC is well reproduced by the fitting curve. The apparent difference in the energy position between the upper and lower peaks demonstrates the existence of a finite Dirac gap, as also corroborated by the extracted bare band dispersions $E_{+1}(k)$ and $E_{-1}(k)$ shown by dashed curves in Figs. \ref{fig:Fig.3}h and \ref{fig:Fig.3}i. This suggests that the energy gap opens in the original bare band, and further indicates that the experimental suppression of spectral weight at the Dirac point cannot be understood by assuming the strongly $\omega$-dependent self-energy effect for the gapless Dirac cone, distinct from the case of graphene on SiC \cite{bostwick2010observation}.

One might expect that the application of Bayesian analysis to a single experimental EDC would be sufficient for concluding a finite Dirac gap in the bare-band dispersion. However, this is not the case because the background shape can be arbitrarily chosen for the sake of just numerically reproducing the single EDC. The analysis of 2D ARPES image itself, in which the background (and matrix element and self-energy as well) is a continuous function of $k$ and $\omega$, is essential for extracting the intrinsic band parameters. Also, the contribution from the $\omega$-dependent self-energy that causes the peak shift and asymmetry in the spectral line shape can never be captured by the analysis of single EDC.

\begin{figure}[tbp]
\begin{center}
\begin{tabular}{c}
\includegraphics[width=8.6cm]{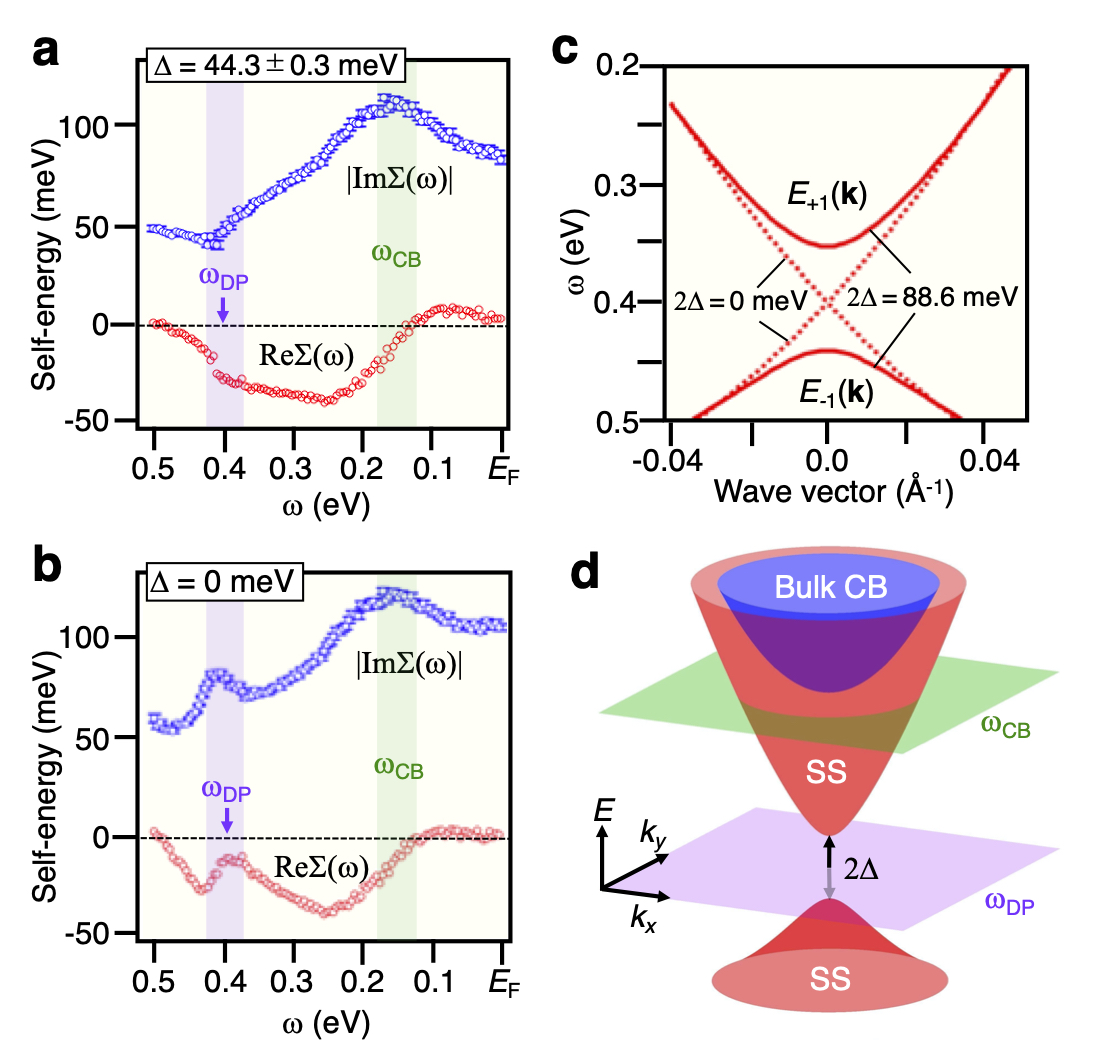}
\end{tabular}
\end{center}
\caption{\textbf{Signature of many-body effects in the Dirac electrons.} {\bf a}, {\bf b} Real (red open circles) and imaginary (blue open circles) parts of electron self-energy ($\mathrm{Re} \Sigma(\omega)$ and $|\mathrm{Im}\Sigma(\omega)|$) extracted from the semiparametric Bayesian analysis for the gapped ($\Delta = 44.3 \pm 0.3$ meV) and gapless ($\Delta = 0$) cases, respectively. Here, $\omega$ and $\Delta$ denote the binding energy and the half-width of band gap, respectively. {\bf c}, Bare-band dispersion for the gapped (red solid curves) and gapless (red dotted curves) Dirac cones used in a and b. {\bf d}, Schematic band diagram of bulk and gapped Dirac-cone bands, together with the energy slices ($\omega_{DP}$ and $\omega_{CB}$) where the self-energy exhibits a characteristic energy dependence. Each error bar of $|\mathrm{Im}\Sigma(\omega)|$ denotes the posterior standard deviation, where each profile of $\mathrm{Re} \Sigma(\omega)$ calculated from each profile of $|\mathrm{Im}\Sigma(\omega)|$ satisfies the Kramers-Kronig relation under assumption of the particle-hole symmetry.}
\label{fig:Fig.4}
\end{figure}

\noindent
\textbf{One-body and many-body characteristics of ${\bf TlBi(S,Se)_2}$.}
Thanks to the extraction of intrinsic Dirac gap through our Bayesian analysis, we can access the intrinsic many-body interactions, which was not possible in the previous studies owing essentially to the uncertainty in determining the bare-band dispersion. We plot in Fig. \ref{fig:Fig.4}a the real and imaginary parts of self-energy, $\mathrm{Re} \Sigma(\omega)$ and $|\mathrm{Im} \Sigma(\omega)|$ simultaneously extracted with $\Delta = 44.3 \pm 0.3$ meV from our Bayesian analysis for the gapped-state model, compared with those obtained from the gapless-state model where $\Delta$ is intentionally fixed to $0$ meV (the invalid case; see also Fig. \ref{fig:Fig.4}c). One can see the overall smooth $\omega$ dependence of both $\mathrm{Re} \Sigma$ and $\mathrm{Im} \Sigma$ for $\Delta = 44.3 \pm 0.3$ meV (Fig. \ref{fig:Fig.4}a) whereas there exists an unusual hump feature around $\omega_{DP}$ in both $\mathrm{Re} \Sigma$ and $\mathrm{Im} \Sigma$ for $\Delta = 0$ (Fig. \ref{fig:Fig.4}b). Such anomaly is unphysical and associated with an artifact originating from the assumption of gapless Dirac-cone state despite a finite Dirac gap. In fact, when the $\Delta$ value is properly incorporated in Fig. \ref{fig:Fig.4}a, such anomaly disappears. One can see from the self-energy plot in Fig. \ref{fig:Fig.4}a that $|\mathrm{Im} \Sigma|$ which reflects the quasiparticle scattering rate (inversely proportional to the quasiparticle lifetime) has a broad maximum at around $\omega \sim 0.15$ eV, whereas it shows a minimum at $\sim 0.4$ eV, around $\omega_{DP}$. The lower scattering rate on approaching $\omega_{DP}$ is reasonable when we consider the available phase space of the Dirac-cone states, because the phase space should monotonically increase on moving away from $\omega_{DP}$ due to the expansion of equi-energy contour in $k$ space, as can be seen from Fig. \ref{fig:Fig.4}d. It is emphasized however that the broad hump seen in $|\mathrm{Im} \Sigma|$ cannot be understood by this argument, requiring the presence of additional scattering channel. As a possible source of this channel, we point out the bulk conduction band which has a bottom at $\omega_{CB} \sim 0.15$ eV (see Fig. \ref{fig:Fig.2}a). When $\omega$ is located in the energy range of bulk conduction band (i.e., $\omega < \omega_{CB}$), the surface-bulk inter-band scattering would take place besides the intra-surface-band scattering, leading to the nonmonotonic behavior of $|\mathrm{Im} \Sigma|$ around $\omega_{CB}$. As shown in Fig. \ref{fig:Fig.4}a, one can also recognize that $|\mathrm{Re} \Sigma|$ becomes maximally $40$ meV, comparable to the size of Dirac gap. This suggests that the influence of self-energy effects cannot be neglected in the band dispersion; in particular, the bare-band dispersion cannot be determined by simply tracing the peak maxima of EDCs.

\section*{Discussion}
The present study sheds light on the fiercely debated origin of Dirac gap in TIs. The Dirac gap of $\mathrm{TlBi}(\mathrm{S}_{1-x} \mathrm{Se}_x)_2$ (as well as those seen in magnetically doped TIs) has been interpreted in terms of many different scenarios standing either on the intrinsically massive Dirac fermions or the massless ones. The former involves the hybridization between surface and interface Dirac cones \cite{zhang2010crossover}, hybridization with impurity bands \cite{sanchez2016nonmagnetic}, local symmetry breaking \cite{tanaka2018influence}, disorder-driven topological phase transition \cite{brahlek2016disorder}, and chemical-inhomogeneity-induced smearing of band inversion \cite{zhang2018topological}. The latter based on the massless Dirac fermions can be associated with the extremely strong coupling with collective modes (including phonon \cite{li2013phonon} and plasmaron \cite{bostwick2010observation}), spin dephasing \cite{qi2019dephasing}, exciton pairing \cite{wang2017excitonic}, and the final-state effect etc. The present study that applies the semiparametric Bayesian modelling to ARPES data suggests that the latter approach is unlikely to be responsible for the Dirac gap. To be more specific, taking into account of the intrinsic gap on the bare band as well as the behavior of self-energy around $\omega_{DP}$ in Fig. \ref{fig:Fig.4}a which can be basically explained in terms of the phase-space argument for the ordinary Dirac electrons, it is suggested that many-body effects such as the electron-electron scattering and the electron-mode coupling, as intensively discussed in strongly correlated systems like high-temperature superconductors, are not responsible for the formation of massive Dirac fermions in $\mathrm{TlBi}(\mathrm{S}_{1-x} \mathrm{Se}_x)_2$. It is thus inferred that the observed gap in $\mathrm{TlBi}(\mathrm{S}_{1-x} \mathrm{Se}_x)_2$ is different from the “gap” seen in the Dirac-cone band of graphene that was suggested to be associated with the many-body interactions \cite{bostwick2010observation}. 

Since the above consideration supports an intrinsically massive Dirac fermion in $\mathrm{TlBi}(\mathrm{S}_{1-x} \mathrm{Se}_x)_2$, it would be useful to compare the present result with the results for magnetic topological insulators where the Dirac gap is expected to open due to the time-reversal-symmetry breaking but not due to the exotic many-body interactions. The magnitude of experimental Dirac gap in the magnetic topological insulators such as $\mathrm{MnBi_2Te_4}$ (e.g., \cite{li2019dirac, chen2019intrinsic, gong2019experimental, otrokov2019prediction, hao2019gapless, chen2019topological}) and topological insulators proximitized with ferromagnets is very small or even undetectable by ARPES, in contrast to the sizable Dirac-gap magnitude of $2\Delta = 88.6$ meV revealed by the Bayesian analysis for $\mathrm{TlBi(S_{0.2}Se_{0.8})_2}$. This result, together with the fact that $\mathrm{TlBi}(\mathrm{S}_{1-x} \mathrm{Se}_x)_2$ shows no magnetic order, suggests that a possibility of local time-reversal-symmetry breaking due to the local magnetic order is ruled out to account for the observed Dirac gap.

The semiparametric Bayesian modelling of ARPES data proposed in this study can be widely applicable to various Dirac-electron systems where the interplay among the Dirac gap, symmetry breaking, and many-body interactions is of interest, as represented by the Dirac-band anomaly in magnetic TIs, axion insulators, and graphene. Also, when the appropriate analytical form of bare-band is established, the Bayesian-based approach would work effectively in a wider variety of systems characterized by the band anomaly occurring in a small energy scale, such as the spin-orbit gap due to the band inversion, the small band splitting associated with the spin-orbit coupling, and the dispersion kink due to the electron-mode coupling.

\section*{Methods}
\noindent
\textbf{Bayes’ formula.}
Throughout our analyses, the chain rule of probability $\mathrm{Pr}(B|A) = \mathrm{Pr}(A|B) \mathrm{Pr}(B) / \mathrm{Pr}(A)$ for random variables $A$ and $B$, called Bayes’ formula, was utilized. For the parameter estimation in the EDC analysis, $B$ corresponds to the set {width, position, intensity} for each peak (parameter set), while $A$ corresponds to the set \{EDC data, the model class (peak number $K$), “temperature”\} (note that the inset to Fig. \ref{fig:Fig.1}c represents this $\mathrm{Pr}(B|A)$, referred to as posterior probability distribution of the parameter set). Under the Bayes’ formula, what one needs to carry out is the modelling of $\mathrm{Pr}(A|B)$ and $\mathrm{Pr}(A)$, called here the likelihood function and prior probability distribution, respectively. Once these models are formulated, their appropriateness can also be evaluated by the Bayes’ formula with the relation $\mathrm{Pr}(A) =\sum \mathrm{Pr}(A|B) \mathrm{Pr}(B)$. For the model selection in the EDC analysis, the relation $\mathrm{Pr}(K | \mathrm{EDC data}) = \mathrm{Pr}(\mathrm{EDC data} | K) \mathrm{Pr}(K) / \mathrm{Pr}(\mathrm{EDC data})$ was utilized, where $\mathrm{Pr}(K | \mathrm{EDC data})$ is referred to as posterior probability distribution of parameter set (Fig. \ref{fig:Fig.1}b). Note that $\mathrm{Pr}(K | \mathrm{EDC data})$ is derived by integrating out “temperature” in $\mathrm{Pr}(A)$ since $\mathrm{Pr}(A) = \mathrm{Pr}(\mathrm{EDC data}, \mathrm{"temperature"} | K) \mathrm{Pr}(K)$ holds.
\\

\noindent
\textbf{Bayesian analysis of EDC.}
In the EDC analysis (Fig. \ref{fig:Fig.1}), the posterior probability distribution for the parameter set was formulated by $p \propto \exp(-n \beta \mathrm{MSE})\phi$, where $n$, $\beta$, $\phi$, and MSE are number of data points constructing EDC, a hyper-parameter (“inverse temperature”), the prior probability distribution, and the mean square error, respectively. The MSE for each $K$ is defined by a difference between EDC data and sum of all Lorentzian functions. The function $\phi$ was set as the continuous uniform distribution whose support is $[0, 0.1]$ (eV) for the peak width, $[-0.5, -0.3]$ (eV) for the peak position, and $[0, 1]$ (a.u.) for the peak intensity.
\\

\noindent
\textbf{Formulation of semiparametric Bayesian analysis.}
In the analysis of 2D ARPES image (Figs. \ref{fig:Fig.3} and \ref{fig:Fig.4}), we assumed that the intensity $Y_{ij}$ of ARPES image at each pixel $(k_i, \omega_j)$ for $i = 1,\cdots, m$ and $j = 1, \cdots, n$ is given by $Y_{ij}=I(k_i,\omega_j;w_0,E_k)+\xi_{ij}$, where $I$ is the intensity function defined by an equation in Fig. \ref{fig:Fig.2}d, $E_k$ analytical form of bare-band dispersion, and $w_0$ the set of other elements including parameters of bare-band dispersion, the self-energy, the matrix elements, and the “background”. The random variable $\xi_{ij}$ is observation noise subject to the Gaussian distribution whose mean and variance are $0$ and $\beta_0^{-1} > 0$, respectively. In other words, $Y_{ij}$ is assumed to be subject to the conditional probability density function
\begin{align}
& p\left(y_{i j} \mid k_{i}, \omega_{j} ; w, E_{s}, \beta\right) \notag \\
&:=\sqrt{\frac{\beta}{2 \pi}} \exp \left(-\frac{\beta}{2}\left(Y_{i j}-I\left(k_{i}, \omega_{j} ; w, E_{s}\right)\right)^{2}\right)    
\end{align}
with $w = w_0$, $E_s = E_k$, and $\beta = \beta_0$. Since $w_0$, $E_k$, and $\beta_0$ are unknown in practice and should be estimated, we treat them as random elements $w$, $E_s$, and $\beta$ subject to the posterior probability density function
\begin{align}
&p\left(w \mid D^{m n}, E_{s}, \beta\right) \notag \\
&=\frac{\phi(w)}{Z\left(E_{s}, \beta\right)} \prod_{i=1}^{m} \prod_{j=1}^{n} p\left(Y_{i j} \mid k_{i}, \omega_{j} ; w, E_{s}, \beta\right)
\end{align}
where $D^{mn} = \{Y_{ij}, k_i, \omega_j\}$ is a data set of ARPES image, $\phi(w)$ an arbitrary prior probability density function, and $Z(E_s,\beta):=p(\{y_{ij}\}|\{k_i\},\{\omega_j\},E_s,\beta)$ the partition function. Note that $w$ consists of $2n+3m+4$ (or $2n+3m+3$) scalar values for the gapped (or gapless) state: the binding energy at the Dirac point $\omega_{DP}$, the band asymmetry $\alpha$, a band parameter $\gamma$, the half-width of band gap $\Delta$ ($\Delta = 0$ for the gapless state), the imaginary part of self-energy $\mathrm{Im}\Sigma(\omega_j)\}$, the matrix elements $\{M_{+1}(k_i), M_{-1}(k_i), M_B(k_i)\}$, and the “background” $\{B(\omega_j)\}$. The function $\phi$ was set as follows: the exponential distribution whose mean is $10$ (eV$\cdot\mathrm{\AA}^2$) for $\alpha$, $4$ (eV$\cdot \mathrm{\AA}$) for $\gamma$, $0.1$ (eV) for $|\mathrm{Im}\Sigma|$, and $0.2$ (a.u.) for $M_s(k)$, the continuous uniform distribution whose support is $[0, 0.5]$ (eV) for $\omega_{DP}$, $[0, 0.25]$ (eV) for $\Delta$, $[0, 1]$ (a.u.) for $M_B(k)$, and $[0, 10]$ (a.u.) for $B(\omega)$.

We should also mention that $p(w | D^{mn},E_s,\beta) \propto f(w;E_s,\beta)$ holds for the function
\begin{align}
f\left(w ; E_{s}, \beta\right):=\phi(w) \exp \left(-\frac{n m \beta}{2} \mathrm{MSE}\left(w ; E_{s}\right)\right),
\end{align}
with the mean square error function
\begin{align}
\mathrm{MSE}\left(w ; E_{s}\right):=\frac{1}{n m} \sum_{i=1}^{m} \sum_{j=1}^{n}\left(Y_{i j}-I\left(k_{i}, \omega_{j} ; w, E_{s}\right)\right)^{2}.
\end{align}
The Bayesian analysis treats the statistical ensemble of $w$ subject to $p(w|D^{mn},E_s,\beta)$ as an extension of the least-squares method. The mean and standard deviation of $p(w|D^{mn},E_s,\beta)$ is respectively adopted as estimator and its error bar. We also estimated $E_s$ and $\beta$ by treating them as random elements subject to the conditional probability distribution function
\begin{align}
p\left(E_{s}, \beta \mid D^{m n}\right)=\frac{Z\left(E_{s}, \beta\right)}{\sum_{\left\{E_{s}\right\}} \int Z\left(E_{s}, \beta\right) d \beta},
\end{align}
where $\{E_s\}$ is a collection of candidate forms of $E_s$. Note that this equation is derived from Bayes’ formula such that $p(E_s, \beta)$ is an uniform distribution. Especially, $E_s$ and $\beta$ that maximize $p(E_s,\beta | D^{mn})$ are adopted as estimators. This type of estimators is known as the empirical Bayes estimator \cite{mackay1992bayesian, bishop2006pattern, tokuda2017simultaneous}. We also quantify the uncertainty of each $E_s$ by the marginal probability
\begin{align}
p\left(E_{s} \mid D^{m n}\right)=\int p\left(E_{s}, \beta \mid D^{m n}\right) d \beta,
\end{align}
as shown in the inset of Fig. \ref{fig:Fig.3}a (see also Fig. \ref{fig:Fig.1}b).
\\

\noindent
\textbf{Algorithm of semiparametric Bayesian analysis.}
The computation of $p(w|D^{mn},E_s,\beta)$ was performed by the exchange Monte Carlo method \cite{geyer1991markov, hukushima1996exchange} (see also Table \ref{tab:Table.S1}), where $\beta$ is discretized as 128 points consisting of 0 and 127 logarithmically spaced points in the interval $[1.5 \times 10^{-10}, 1.5 \times 10^2]$. The total Monte Carlo sweeps were $10,000$ after the burn-in, where the obtained sequence $\{w_l^t\}$ for $t = 1, \cdots, 10,000$ and $l = 1, \cdots, 128$ is regarded as a statistical ensemble of $w$ subject to $p(w|D^{mn},E_s,\beta_l)$. Figures 1c and 3a-3d are the density scatter plots of $\{w_l^t\}$ at $\beta$ that maximize $Z(E_s,\beta)$.
We also calculate $p(E_s,\beta|D^{mn})$ via the bridge sampling \cite{meng1996simulating, gelman1998simulating}, as shown by
\begin{align}
Z\left(E_{s}, \beta_{l}\right) &= \prod_{l^{\prime}=1}^{l-1} \frac{Z\left(E_{s}, \beta_{l^{\prime}+1}\right)}{Z\left(E_{s}, \beta_{l^{\prime}}\right)} \notag \\
&=\prod_{l^{\prime}=1}^{l-1}\left\langle\exp \left(-\frac{n m}{2}\left(\beta_{l^{\prime}+1}-\beta_{l^{\prime}}\right) \mathrm{MSE}\left(w ; E_{s}\right)\right)\right\rangle_{\beta_{l^{\prime}}},
\end{align}
where $\langle Q\rangle_\beta$ denotes the average of an arbitrary quantity $Q$ over $p(w|D^{mn},E_s,\beta)$ and is approximated by sample mean of obtained sequence $\{Q_l^t\}$.

\section*{Data availability}
The data and information within this paper are available from the corresponding authors upon request.

\section*{Code availability}
The computer code to generate the results are available from the corresponding authors upon request.

\bibliography{Tokuda_main.bbl}

\begin{acknowledgments}
This work was supported by JST-CREST (no. JPMJCR18T1), Grant-in-Aid for Scientific Research on Innovative Areas “Topological Materials Science” (JSPS KAKENHI Grant number JP15H05853), Grant-in-Aid for Scientific Research on Innovative Areas “Discrete Geometric Analysis for Materials Design” (JSPS KAKENHI Grant number JP18H04472), and Grant-in-Aid for Early-Career Scientists (JSPS KAKENHI Grant number 20K19889). The work in Cologne was funded by the Deutsche Forschungsgemeinschaft (German Research Foundation) - Project number 277146847 - CRC 1238 (Subproject A04).
\end{acknowledgments}

\section*{AUTHOR CONTRIBUTIONS}
The work was planned and proceeded by discussion among S.T., S.S, T.N. T.T. and T.S. S.T. carried out the semiparametric Bayesian analysis. K.S. and Y.A. carried out the sample growth. S.S. and T.S performed the ARPES measurements. S.T. and T.S. finalized the manuscript with inputs from all the authors.

\section*{COMPETING INTERESTS}
The authors declare no competing interests.

\section*{Supplementary Note 1: Demonstration of the semiparametric Bayesian analysis}
To validate our methodology, we conduct a demonstration by using synthetic data that the electronic structure is predefined as a ground truth. Two types of synthetic images $D^{mn}$ as simulated ARPES intensities are shown in Fig. \ref{fig:Fig.S1}: gapless (Fig. \ref{fig:Fig.S1}a) and gapped (Fig. \ref{fig:Fig.S1}d) Dirac cones. Both two images consist of $250$ intensities $\{Y_{ij}\}$ numerically generated from $p(y_{ij}|k_i,\omega_j;w,E_s,\beta)$ with $m = 50$ linearly spaced points $\{k_i\}$ in the interval $[-0.05, 0.05] \mathrm{\AA}^{-1}$ and $n = 50$ linearly spaced points $\{\omega_j\}$ in the interval $[0, 0.5]$ eV. The predefined elements are as follows: $\omega_{DP} = 0.25$ eV, $\alpha=0$ eV$\cdot \mathrm{\AA}^2$, $\gamma = 5$ eV$\cdot \mathrm{\AA}$, $\Delta = 0$ eV (gapless) or $0.05$ eV (gapped), $\mathrm{Im}\Sigma(\omega) = -0.05$ eV, $M_{+1}(k) = M_{-1}(k) = 0.1$, $M_B(k) = 0$, $B(\omega) = 0$, and $\beta = 10^2$, where the signal-to-noise ratio $\mathrm{max}(A_s M_s) \sqrt{\beta}=20$. We also assume that the temperature is absolute zero. In the semiparametric Bayesian analysis, $\phi(w)$ is set as follows: the exponential distribution whose mean is $0.01$ (eV$\cdot \mathrm{\AA}^2$) for $\alpha$, $5$ (eV$\cdot \mathrm{\AA}$) for $\gamma$, $0.05$ (eV) for $|\mathrm{Im}\Sigma|$, and $0.1$ (a.u.) for $M_s(k)$, the continuous uniform distribution whose support is $[0, 0.5]$ (eV) for $\omega_{DP}$, $[0, 0.25]$ (eV) for $\Delta$, $[0, 1]$ (a.u.) for $M_B(k)$, and $[0, 1]$ (a.u.) for $B(\omega)$. The setup of algorithm, namely the Monte Carlo simulation, is the same as Methods section in the main text.
A side-by-side comparison of the synthetic ARPES image for the gapless Dirac-cone band (Fig. \ref{fig:Fig.S1}a) and its reproduction by $I(k, \omega)$ obtained from the semiparametric Bayesian analysis (Fig. \ref{fig:Fig.S1}b) signify the almost identical intensity distribution except for the noise components. Such a good matching is highlighted by the consistency between the true and estimated bare-band dispersions in Fig. \ref{fig:Fig.S1}c. One can further confirm whether the Dirac-cone state is gapless or not by comparing $Z(E_s, \beta)$ of each $E_s$ for gapless ($\Delta = 0$) and gapped ($\Delta > 0$) models. The maximum of $Z(E_s, \beta)$, namely the maximum of $p(E_s, \beta| D^{mn})$, is at that $E_s$ is gapless and in the vicinity of $\beta = 10^2$ (Fig. \ref{fig:Fig.S1}g); The empirical Bayes estimator is consistent with ground truth. One can immediately recognize in the inset of Fig. \ref{fig:Fig.S1}g that $p(E_s | D^{mn})$ for $\Delta > 0$ is negligibly small ($0.11$\%) as opposed to the case for $\Delta = 0$ ($99.89$\%). The same examination for the synthetic ARPES image for the gapped Dirac-cone band also shows the validity of our methodology, as shown in Figs. \ref{fig:Fig.S1}d-\ref{fig:Fig.S1}f, and \ref{fig:Fig.S1}h.

\section*{Supplementary Note 2: Extracted spectral components from the semiparametric Bayesian analysis}
Thorough the semiparametric fittings of 2D ARPES intensity for $\mathrm{TlBi}(\mathrm{S}_{1-x} \mathrm{Se}_x)_2$ ($x = 0.8$), we have also obtained matrix-element term of photoelectron intensity and background function, besides the spectral function $A_s(k, \omega)$ shown in Figs. 3h and 3i of the main text. We show in Fig. \ref{fig:Fig.S2}a the extracted matrix-element term for the upper and lower Dirac cones, $M_{+1}(k)$ and $M_{-1}(k)$, respectively, in both the gapped and gapless cases. One can immediately recognize unusually strong $k$ dependence for both $M_{+1}(k)$ and $M_{-1}(k)$ around $k = 0$ only for the gapless case. This strongly suggests that the assumption of gapless Dirac cone is unphysical because the matrix-element term is generally a moderate function of $k$ in the narrow $k$ range. This is also consistent with our conclusion drawn from the unusual behavior of self-energy $\Sigma$ in Fig. 4b of the main text. We also found in Figs. \ref{fig:Fig.S2}b and \ref{fig:Fig.S2}c that the matrix-element term for the background $M_B(k)$, and the background function $B(\omega)$, show moderate $k$ and $\omega$ dependences, respectively, and they are not so sensitive to the behavior of the Dirac gap (i.e. gapless vs gapped). This may be reasonable since these terms are associated with the bulk band, but not with the Dirac-cone surface state.

\renewcommand{\thefigure}{S\arabic{figure}}
\setcounter{figure}{0}
\renewcommand{\thetable}{S\arabic{table}}
\setcounter{table}{0}

\begin{figure*}[tbp]
\begin{center}
\begin{tabular}{c}
\includegraphics[width=11.2cm]{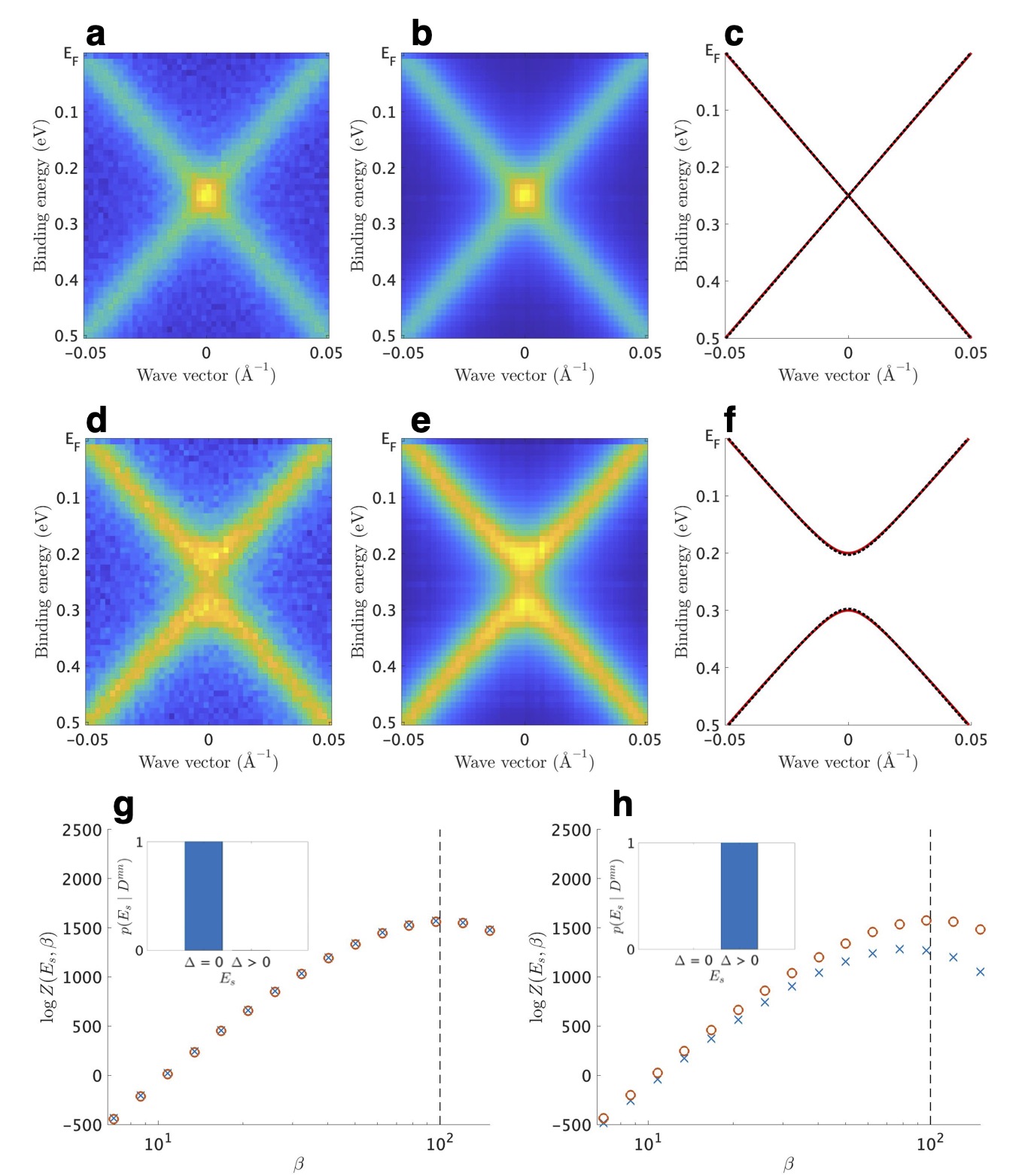}
\end{tabular}
\end{center}
\caption{\textbf{Semiparametric Bayesian analysis for synthetic images with ground truths.} {\bf a}, Intensity plot of synthetic ARPES image for gapless Dirac-cone bands. {\bf b}, Reproduced $I(k, \omega)$ obtained from the semiparametric fitting with gapless Dirac-cone band. {\bf c}, Bare-band dispersion for the gapless Dirac cones in a (red solid line) and b (black dotted line). {\bf d}-{\bf f}, The same as a-c for gapped Dirac-cone bands. {\bf g}, Plot of $Z(E_s, \beta)$ given $D^{mn}$ of a against $\beta$ for the case that $E_s$ is gapless (blue crosses) or gapped (red circles), compared with the ground truth of $\beta$ (black dashed line). Inset shows $p(E_s | D^{mn})$ that supports the existence of gapless Dirac-cone band in a. {\bf h}, The same as g for given $D^{mn}$ of b.}
\label{fig:Fig.S1}
\end{figure*}

\begin{figure*}[h]
\begin{center}
\begin{tabular}{c}
\includegraphics[width=17.2cm]{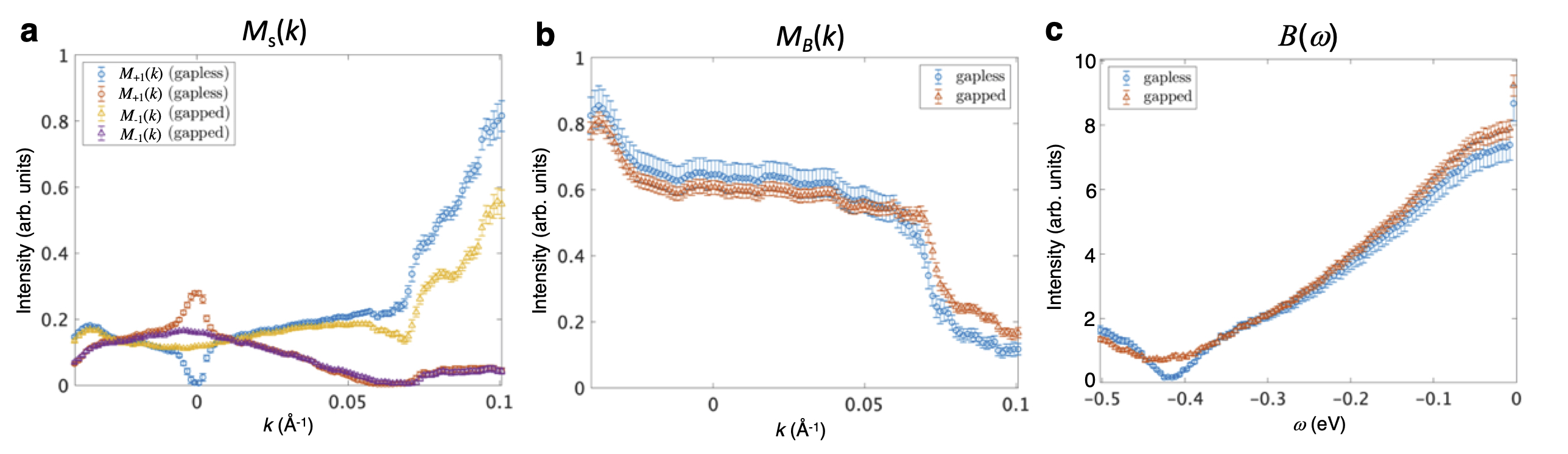}
\end{tabular}
\end{center}
\caption{\textbf{Matrix-element and background terms in the semiparametric Bayesian analysis.} {\bf a}-{\bf c}, Simulated k-dependent matrix-element term of photoelectron intensity for the Dirac-cone state $M_s(k)$ ($s = \pm 1$), that for the background $M_b(k)$, and the $\omega$-dependent background state $B(\omega)$, respectively, for the gapped ($\Delta = 44.3 \pm 0.3$ meV) and gapless ($\Delta = 0$ meV) cases.}
\label{fig:Fig.S2}
\end{figure*}

\begin{table*}[h]
    \centering    
    \caption{\textbf{Basic algorithm of exchange Monte Carlo method.}}
    \begin{tabular}{c l}
        \hline
    (1) & Discretize $\beta \geq 0$ as $0=\beta_1<\beta_2<,\cdots,\beta_L$ in an arbitrary interval $[\beta_1, \beta_L]$ \\
	(2) & Choose an arbitrary initial value $w_l^1 \in W$ for $l=1,\cdots,L$ and an arbitrary \\
	& probability density function $g_l({w_l}'|w_l^t )$ such that $g_l({w_l}'|w_l^t) = g_l(w_l^t|{w_l}')$ \\
	(3) & Sample ${w_l}'$ from $g({w_l}'|w_t)$ for each $l$ \\
	(4) & Calculate the acceptance ratio $r_l=  f({w_l}';E_s,\beta_l)/f(w_l^t;E_s,\beta_l)$  for each $l$ \\
	(5) & Generate a uniform random number $u_l \in [0,1]$ for each $l$ \\
	(6) & Set $w_l^{t+1}={w_l}'$ for each $l$ if $u_l \leq r_l$ \\
	(7) & Set $w_l^{t+1}=w_l^t$ for each $l$ if $u_l > r_l$ \\
	(8) & Calculate the exchange ratio \\
    & $R_l= \exp(\frac{nm}{2} (\beta_l-\beta_{l-1} )(\mathrm{MSE}(w_l^{t+1};E_s)-\mathrm{MSE}(w_{l-1}^{t+1};E_s)))$ for $l=2,\cdots,L$ \\
	(9) & Generate a uniform random number $v_l \in [0,1]$ for each $l$ \\
	(10) & Swap $w_l^{t+1}$ and $w_{l-1}^{t+1}$ for each $l$ if $v_l \leq R_l$ \\
    (11) & Repeat (3)-(10) for $t = 2, 3, \cdots, T$ \\
        \hline
    \end{tabular}
    \label{tab:Table.S1}
\end{table*}

\end{document}